\documentstyle[12pt,epsf]{article}

0

\begin{document}
\baselineskip 7mm

\begin{flushright}
IFT-UAM/CSIC-97-3\\
hep-th/9711074  
\end{flushright}

\begin{center}

{}~\vfill

{\large \bf  Elliptic Singularities, $\theta$-Puzzle and Domain Walls }

\end{center}

\vspace{10 mm}

\begin{center}

{\bf C\'{e}sar G\'{o}mez} 
  
\vspace{7 mm}

{\em Instituto de Matem\'{a}ticas y F\'{\i}sica Fundamental,
CSIC, \protect \\ Serrano 123, 28006 Madrid, Spain}  

\vspace{5 mm}
  
{\em and}
  
\vspace{5 mm}

{\em Instituto de F\'{\i}sica Te\'{o}rica (UAM-CSIC), 
C-XVI, Universidad Aut\'{o}noma de Madrid, Cantoblanco 28049,
Madrid, Spain}

\vspace{5mm}

\end{center}      

\vspace{15mm}
   
\begin{abstract}

We study $N=1$ four dimensional gluodynamics in the context of
M-theory compactifications on elliptically fibered Calabi-Yau
fourfolds. Gaugino condensates, $\theta$-dependence, Witten index
and domain walls are considered for singularities of type
$\hat{A}_{n-1}$ and $\hat{D}_{n+4}$. It is shown how the topology
of intersections among the irreducible components defining the
singular elliptic fiber, determine the entanglement of vacua and
the appareance of domain walls.

\end{abstract}

\pagebreak

\section{Introduction and Summary}

$N=1$ four dimensional gauge theories can be obtained by means of
$F$-theory compactifications \cite{Wsp,svw,dgw,gm} on elliptically fibered
Calabi Yau fourfolds $X$ :
\begin{equation}
E \rightarrow X \rightarrow B,
\label{1}
\end{equation}
or equivalently by $M$-theory compactifications on $X$ in the limit
$\hbox {Vol }(E)=0$ . Assuming Kodaira classification \cite{Kodaira}
of singular fibers extends to the case of elliptically fibered
Calabi-Yau fourfolds and working locally, in the spirit of geometric
engineering \cite{kkv,KV}, we can get $N=1$ four dimensional gauge
theories with $ADE$ type of gauge groups. In this paper we will
consider, from that point of view, some problems concerning $N=1$
super gluodynamics, in particular gaugino condensates, the
$\theta$-puzzle \cite{Smil}, the Witten index \cite{Wind} and domain
walls \cite{ds,kss,sv}. These issues has been recently considered, in
the context of intersecting branes in reference \cite{Wbqcd}.

Chiral symmetry breaking in $N=1$ pure gluodynamics, i. e., non
vanishing condensates $<\lambda \lambda>$ for gluino bilinears, has
been derived long time ago using different methods. In $SU(n)$ pure
super Yang-Mills, instantons contribute to condensates $<\lambda
\lambda(x_{1}) \ldots \lambda \lambda(x_{n})>$ and they are, after
integration over the size of the instanton, independent of the
positions $x_{1},x_{2},\ldots,x_{n}$ \cite{NSVZ1,NSVZ2,A}.  Through
cluster decomposition we can get a set of values for
$<\lambda\lambda>$,
\begin{equation}
<\lambda \lambda>_j = C \Lambda^3 e^{2 \pi ij/n} e^{i \theta/n}
\label{2}
\end{equation}
with $j=0, \ldots, n-1$, $C$ some constant, $\Lambda$ the $SU(n)$ $N=1$ scale and
$\theta$ the $\theta$-vacua parameter. A similar result
\cite{CG} can be directly derived, for $SU(n)$ gauge groups,
using `t Hooft's torons \cite{tHcmp}. In both cases 
this way of computing the
condensates is doubtful. Namely, using cluster, we must implicetely consider
the contribution of instantons of arbitrarily large size, which
is certainly beyond the regime where semiclassical analysis is
reliable. The same happens in the toron computation, where an
infinite volume limit must be performed before reaching the
result (\ref{2}). A different approach to the computation of
$<\lambda\lambda>$ is in the weak coupling regime \cite{SV}
where the holomorphy dependence on an auxiliary mass is used.
These two procedures do not coincide on the value of the constant
in (\ref{2}). Recently a possible way out to this
puzzle has been proppossed in reference \cite{KS} where the
existence of an extra vacua without chiral symmetry breaking is
suggested. Leaving for a while these issues the result (\ref{2})
contains already some interesting peculiarities. If the different
values for $<\lambda\lambda>_{i}$ are identified with different
vacua, the change $\theta \rightarrow \theta + 2 \pi$ transforms the vacuum $i$ into the
vacuum $i+1$. Moreover the number of $<\lambda\lambda>$ values
coincides, for $SU(n)$ with the Witten index $\hbox {tr }(-1)^F$, but this is not
the case for $O(N)$ groups where the number of $<\lambda\lambda>$
values is $N-2$ while the Witten index is $\left[ \frac {N}{2} \right] +1 $.
  
Since non vanishing values for $<\lambda\lambda>$ are connected
with a non trivial superpotential for $N=1$ gluodynamics
\cite{IS}, and taking into account that vertical instantons
\cite{Wsp} in M-theory survive in the $N=1$ four dimensional
limit, it is natural looking for a superpotential that agrees
with (\ref{2}), directly in M-theory. A first step in this
direction was already taken in reference \cite{KV}. In this
approach the geometry of Kodaira singularities is crucial.
For an elliptically fibered Calabi-Yau fourfold, M-theory
instantons \cite{Wsp} are defined as divisors,  i. e.,  six cycles,
with holomorphic Euler characteristic $\chi=1$. If the instanton is
vertical, then we can safely use it as a contribution to the four
dimensional superpotential. In the Kodaira classification
\cite{Kodaira}, the singular fibers are defined by 2-cycles ${\cal C}$ of
the type:
\begin{equation}
{\cal C} = \sum_{i} n_i \Theta_i,
\label{3}
\end{equation}
with $\Theta_i^2=-2$, non singular rational curves, $n_{i}$ integer numbers
and the intersection matrix $\Lambda_{ij} = (\Theta_i . \Theta_j)$ one to one related to some
affine Dynkin diagram. Moreover, the self intersection $({\cal C}^{2})$ is
equal to zero, which is characteristic of elliptic
singularities. In a Calabi-Yau fourfold $X$ the singular locus
will be a four dimensional manifold $C$ and we will assume that
the singular fiber is trivially fibered over $C$. In addition, we
will think of $C$ as an Enriques surface. With these assumptions,
it is easy to get M-theory instantons $D_{i}$ associated with
the $\Theta_i$ in (\ref{3}) \cite{KV} all of them with $\chi=1$, and
therefore with two fermionic zero modes. Following \cite{Wsp} we
associate with each of these divisors a scalar field $\phi_{D_i}$
transforming under chiral $U(1)$ as:
\begin{equation}
\phi_{D_i} \rightarrow \phi_{D_i} + \alpha.
\label{4}
\end{equation}
The topological sum defined as $\sum n_i \phi_{D_i}$ transforms as $\sum n_i \phi_{D_i} 
\rightarrow \sum n_i \phi_{D_i} + \hbox {Cox }\alpha $, where
$\hbox {Cox}$ is the Coxeter number of the diagram defined as the total
number of irreducible components. This is the transformation rule
of the $\theta$ parameter in $N=1$ super gluodynamics as derived
from the $U(1)$ axial anomaly. Thus we define the $\theta$
parameter as the topological sum $\sum n_i \phi_{D_i} \hbox {  mod} (2 \pi)$. For the $SU(n)$ case
corresponding to $\hat{A}_{n-1}$ singularities, we have under ${\bf Z}_n$
transformations, $ \sum_{i=0}^{n-1} \phi_{D_i} \rightarrow \sum_{i=0}^{n-1} \phi_{D_i} \hbox {mod} (2 \pi)$, 
which can be solved by $\phi_{D_j} = \frac {2 \pi j}{n} + \frac {\theta}{n} $, with
$\theta \in [0, 2 \pi]$ undetermined. This corresponds to the geometry of $\hat{A}_{n-1}$
Kodaira singularities, where under chiral ${\bf Z}_n$ the ireducible
components transform cyclically as $\Theta_ i \rightarrow \Theta_{i+1}$.

By standard M-theory instanton computations \cite{Wsp},
we can associate with each divisor $D_{j}$, such that $\chi(D_j)=1$ a
gaugino condensate $<\lambda\lambda>_{\theta}^j$. From (\ref{4}), and the
definition of $\theta$ as a topological sum, we easily get $< \lambda \lambda>_{\theta}^j = 
<\lambda' \lambda'>_{\theta+ \alpha n}^j$, for $\lambda'=e^{i \alpha/2} \lambda$ (notice
 that in four
dimensions $\theta$ play a similar role to the three dimensional 
Goldstone boson \cite{AHW} which in M-theory on Calabi-Yau
fourfolds can be identified with the scalar field $\phi_{D_i}$).
In summary, 
we start
with divisors $D_{j}$, with $\chi(D_j)=1$ in an elliptically fibered
Calabi-Yau fourfold. By purely geometrical arguments \cite{Wsp}, we
derive the transformation rule (\ref{4}). Next, we consider a
singular fiber of Kodaira type $ADE$ as defined by a cycle of the
type (\ref{3}). Then the topological sum $\sum n_i \phi_{D_i}$ reproduces the
chiral transformations of $\theta$ dictated by the $U(1)$ axial
anomaly of $ADE$ $N=1$ four dimensional gauge theories. At this
point, the reader can wonder where four dimensions enter the
argument. In the three dimensional limit defined by $Vol(E)=\infty$
the singularity, as we will discuss later, becomes rational, of
type $A_{n-1}$, with $n-1$ irreducible components, and the topological
sum does not possess any clear physical meaning.

The $\theta$-puzzle appears as a consequence of the following
fact. If the $\theta$ angle (we reduce here the discussion to
$SU(n)$) is defined as the topological sum $ \sum_{i=0}^{n-1} \phi_{D_i}$, then it is
tempting to think of $\theta$ as the scalar field $\phi_{{\cal D}}$, associated
to the six cycle ${\cal D}$, defined by trivially fibering the
$\hat{A}_{n-1}$ cycle ${\cal C}=\sum_{i=0}^{n-1} \Theta_i $over the 
singular locus $C$. Then the transformation rule under
chiral $U(1)$, $\theta \rightarrow \theta + n \alpha$ will mean that $\chi({\cal D})=n$. This is the
mathematical way to translate the physical argument relating
$\theta$ to instanton tunneling and instantons to condensates of
the type $<\lambda \lambda(x_1) \ldots \lambda \lambda(x_n)>$. Where 
is the loophole of this way of thinking?.
The loophole is mathematically very clear, namely the holomorphic
Euler characteristic of the $6$-cycle ${\cal D}$ is
zero, and not $n$, and therefore $\theta$ is not the field
$\phi{{\cal D}}$.
The intuitive way to understand it, is simply observing that ${\cal C}$
is not just the union of $n$ disconnected components $\Theta_i$, but 
a very specific set of intersections that, by the Grothendieck-Riemann-Roch
theorem \cite{Fulton}, leads to $\chi({\cal D})=0$.

What are the consequences of $\chi({\cal D})=0$ from a physical point of view?.
Intuitively we can think of $\chi({\cal D})=0$ as reflecting a complete soaking
up of the fermionic zero modes associated to the components
$\Theta_i$. This approach allows us to associate to each intersection $(\Theta_i . \Theta_{i+1})$
a term soaking up two
fermionic zero modes, of the same order of magnitude as the
$<\lambda \lambda>$ condensate. These intersection terms are
perfect candidates for domain walls whose existence is the
physical way to solve the $\theta$-puzzle. In other words, $\chi({\cal D})=0$
implies the entanglement of the different vacua, entanglement that is physically 
mediated by the intersection terms. The extension of this entanglemnet 
mechanism to $N=0$ supersymmetry would be extremely interesting for 
explaining the $\theta$ dependence in the Witten-Veneziano 
formula \cite{Wi,Ve}. 
  
Thinking in terms of
branes in M-theory, instantons are interpreted \cite{Wsp} as 
fivebranes wrapped on six cycles $D_{j}$; thus, the intersection term
should be interpreted as a five brane wrapped on $C \times (\Theta_i . \Theta_{i+1})$, which
produces, in four dimensions, a wall of the appropiate tension,
interpolating the $i$ and $i+1$ vacua. Moreover, the two fermionic
zero modes associated to the intersection term seems to be the
right $BPS$ signal of the domain wall. The extension of the
previous arguments to $O(N)$ gauge groups, i. e., to $\hat{D}$ type singularities, will
present some extra difficulties that we will discuss. 

In summary the solution to the $\theta$-puzzle and the existence
of domain walls are related to the topology of intersections of
the irreducible components defining singular elliptic fibers.

\section{Local Models for Elliptic Fibrations.}
\label{sec:elliptic}

Let $V$ be an elliptic fibration,
\begin{equation}
\Phi: V \rightarrow \Delta,
\label{eq:r1}
\end{equation}
with $\Delta$ an algebraic curve, and $\Phi^{-1}(a)$, with $a$
any point in $\Delta$, an elliptic curve. Let us denote
$\{a_{\rho}\}$ the finite set of points in $\Delta$ such that
$\Phi^{-1}(a_{\rho})={\cal C}_{\rho}$ is a singular fiber. Each singular
fiber ${\cal C}_{\rho}$ can be written as
\begin{equation}
{\cal C}_{\rho} = \sum_i n_{i \rho} \Theta_{i \rho},
\label{eq:r2}
\end{equation}
where $\Theta_{i \rho}$ are non singular rational curves, with
$\Theta_{i \rho}^2 =-2$, and $n_{i \rho}$ are integer numbers.
Different types of singularities are characterized by
(\ref{eq:r2}) and the intersection matrix $(\Theta_{i \rho} .
\Theta_{j \rho})$. All different types of Kodaira singularities
satisfy the relation
\begin{equation}
{\cal C}_{\rho}^2=0.
\label{eq:r3}
\end{equation}
  
Let $\tau(u)$ be the elliptic modulus of the elliptic fiber at
the point $u \in \Delta$. For each path $\alpha$ in
$\Pi_1(\Delta')$, with $\Delta'= \Delta- \{a_{\rho}\}$, we can
define a monodromy transformation $S_{\alpha}$, in $Sl(2,{\bf
Z})$, acting on $\tau(u)$ as follows:
\begin{equation}
S_{\alpha} \tau(u) = \frac {a_{\alpha} \tau(u) +
b_{\alpha}}{c_{\alpha} \tau(u) + d_{\alpha}}.
\label{eq:r4}
\end{equation}
Each type of Kodaira singularity is characterized by a particular
monodromy matrix. 
  
In order to define an elliptic fibration \cite{Kodaira}, the starting point will
be an algebraic curve $\Delta$, that we will take, for
simplicity, to be of genus zero, and a meromorphic function
${\cal J}(u)$ on $\Delta$. Let us assume ${\cal J}(u) \neq
0,1,\infty$ on $\Delta' =\Delta-\{a_{\rho}\}$. Then, there
exists multivalued holomorphic function $\tau(u)$, with $\hbox
{Im } \tau(u)>0$, satisfying ${\cal J}(u) = j(\tau(u))$, with $j$
the elliptic modular $j$-function on the upper half plane. As
above, for each $\alpha \in \Pi_1(\Delta')$ we define a monodromy
matrix $S_{\alpha}$, acting on $\tau(u)$ in the form defined by
(\ref{eq:r4}). Associated to these data we will define an elliptic 
fibration, (\ref{eq:r1}). In order to do that, let us first define 
the universal covering $\tilde{\Delta}'$, of $\Delta'$, and let
us identify the covering transformations of $\tilde{\Delta}'$
over $\Delta'$, with the elements in $\Pi_1(\Delta')$. Denoting
by $\tilde{u}$ a point in $\tilde{\Delta}'$, we define, for each
$\alpha \in \Pi_1(\Delta')$, the covering transformation
$\tilde{u} \rightarrow \alpha \tilde{u}$, by
\begin{equation}
\tau(\alpha \tilde{u}) = S_{\alpha} \tau(\tilde{u});
\label{eq:r5}
\end{equation}
in other words, we consider $\tau$ as a single valued holomorphic
function on $\tilde{\Delta}'$. Using (\ref{eq:r4}), we define
\begin{equation}
f_{\alpha}(\tilde{u}) = (c_{\alpha} \tau(\tilde{u}) +
d_{\alpha})^{-1}.
\label{eq:r6}
\end{equation}
Next, we define the product $\tilde{\Delta}' \times {\bf C}$ and,
for each $(\alpha,n_1,n_2)$, with $\alpha \in \Pi_1(\Delta')$,
and $n_1,n_2$ integers, the automorphism
\begin{equation}
g(\alpha,n_1,n_2) : (\tilde{u},\lambda) \rightarrow (\alpha
\tilde{u}, f_{\alpha} (\tilde{u}) (\lambda+n_1 \tau
(\tilde{u})+n_2)).
\label{eq:r7}
\end{equation}
Denoting by ${\cal G}$ the group of automorphisms (\ref{eq:r7}),
we define the quotient space
\begin{equation}
B' \equiv (\tilde{\Delta}' \times {\cal C}) / {\cal G}.
\label{eq:r8}
\end{equation}
This is a non singular surface, since $g$, as defined by
(\ref{eq:r7}), has no fixed points in $\tilde{\Delta}'$. From
(\ref{eq:r7}) and (\ref{eq:r8}), it is clear that $B'$ is an
elliptic fibration on $\Delta'$, with fiber elliptic curves of
elliptic modulus $\tau(u)$. Thus, by the previous construction, we
have defined the elliptic fibration away from the singular points
$a_{\rho}$. 
  
Let us denote $E_{\rho}$ a local neighbourhood of the point
$a_{\rho}$, with local coordinate $t$, and such that
$t(a_{\rho})=0$. Let $S_{\rho}$ be the monodromy associated with
a small circle around $a_{\rho}$. By ${\cal U}_{\rho}$ we will denote
the universal covering of $E_{\rho}'=E_{\rho}-a_{\rho}$, with
coordinate $\rho$ defined by
\begin{equation}
\rho= \frac {1}{2 \pi i} \log t.
\label{eq:r9}
\end{equation}
The analog of (\ref{eq:r5}) will be
\begin{equation}
\tau(\rho+1) = S_{\rho} \tau(\rho).
\label{eq:r10}
\end{equation}
If we go around the points $a_{\rho}$, $k$ times, we should act
with $S_{\rho}^k$; hence, we parametrize each path by the winding
number $k$. The group of automorphisms (\ref{eq:r7}), reduced to
small closed paths around $a_{\rho}$, becomes
\begin{equation}
g(k,n_1,n_2)(\rho,\lambda) =
(\rho+k,f_k(\rho)[\lambda+n_1\tau(\rho)+n_2]).
\label{eq:r11}
\end{equation}
Denoting by ${\cal G}_{\rho}$ the group (\ref{eq:r11}), we define
the elliptic fibration around $a_{\rho}$ as
\begin{equation}
({\cal U}_{\rho} \times {\bf C})/{\cal G}_{\rho}.
\label{eq:r12}
\end{equation}
  
Next, we will extend the elliptic fibration to the singular point
$a_{\rho}$. We can consider two different cases, depending on the
finite or infinite order of $S_{\rho}$.

\subsection{Singularities of Type $\hat{D}_4$: ${\bf Z}_2$
Orbifolds.}

Let us assume $S_{\rho}$ is of finite order,
\begin{equation}
(S_{\rho})^m = {\bf 1}_d.
\label{eq:r13}
\end{equation}
In this case, we can extend (\ref{eq:r12}) to the singular
points, simply defining a new variable $\sigma$ as
\begin{equation}
\sigma^m = t.
\label{eq:r14}
\end{equation}
Let us denote $D$ a local neighbourhood in the $\sigma$-plane of
the point $\sigma=0$, and define the group $G_D$ of automorphisms
\begin{equation}
g(n_1,n_2):(\sigma,\lambda) = (\sigma,\lambda+n_1
\tau(\sigma)+n_2),
\label{eq:r15}
\end{equation}
and the space
\begin{equation}
F=(D \times {\bf C})/G_D.
\label{eq:r16}
\end{equation}
Obviously, $F$ defines an elliptic fibration over $D$, with fiber
$F_{\sigma}$ at each point $\sigma \in D$, an elliptic curve of
modulus $\tau(\sigma)$. From (\ref{eq:r13}) and (\ref{eq:r6}), it
follows that
\begin{equation}
f_k(\sigma) = 1,
\label{eq:r17}
\end{equation}
with $k=O(m)$. Thus, we can define a normal subgroup ${\cal N}$
of ${\cal G}_{\rho}$ as the set of transformations
(\ref{eq:r11}):
\begin{equation}
g(k,n_1,n_2):(\rho,\lambda) \rightarrow (\rho+k,\lambda+n_1
\tau(\rho)+n_2).
\label{eq:r18}
\end{equation}
Comparing now (\ref{eq:r15}) and (\ref{eq:r18}), we get
\begin{equation}
({\cal U}_{\rho} \times {\bf C})/{\cal N} = (D' \times {\bf
C})/G_D \equiv F-F_0.
\label{eq:r19}
\end{equation}
Using (\ref{eq:r18}) and (\ref{eq:r11}) we get
\begin{equation}
{\cal C} = {\cal G}/{\cal N},
\label{eq:r20}
\end{equation}
with ${\cal C}$ the cyclic group of order $m$, defined by
\begin{equation}
g_k:(\sigma,\lambda) \rightarrow (e^{2 \pi i k/m} \sigma,
f_k(\sigma) \lambda).
\label{eq:r21}
\end{equation}
From (\ref{eq:r20}) and (\ref{eq:r19}), we get the desired
extension to $a_{\rho}$, namely
\begin{equation}
F/{\cal C} = ({\cal U}_{\rho} \times {\bf C})/{\cal G}_{\rho}
\cup F_0/{\cal C}.
\label{eq:r22}
\end{equation}
Thus, the elliptic fibration extended to $a_{\rho}$, in case
$S_{\rho}$ is of finite order, is defined by $F/{\cal C}$. Now,
$F/{\cal C}$ can have singular points that we can regularize. The
simplest example corresponds to
\begin{equation}
S_{\rho} = \left( \begin{array}{cc} -1 & 0 \\ 0 & -1 \end{array}
\right),
\label{eq:r23}
\end{equation}
i. e., a parity transformation. In this case, the order is $m=2$,
and we define $\sigma$ by $\sigma^2 \equiv t$. The cyclic 
group (\ref{eq:r21}) in this case simply becomes
\begin{equation}
(\sigma, \lambda) \rightarrow (- \sigma, -\lambda),
\label{eq:r24}
\end{equation}
since from (\ref{eq:r23}) and (\ref{eq:r6}) we get $f_1=-1$. At the point 
$\sigma=0$ we have four fixed points, the standard ${\bf Z}_2$ 
orbifold points,
\begin{equation}
(0,\frac {a}{2}\tau(0)+ \frac {b}{2}),
\label{eq:r25}
\end{equation}
with $a,b=0,1$. The resolution of these four singular points will produce 
four irreducible components, $\Theta^1, \ldots, \Theta^4$. In addition, 
we have the irreducible component $\Theta_0$, defined by the curve itself 
at $\sigma=0$. Using the relation $\sigma^2=t$, we get the $\hat{D}_4$ cycle, 
\begin{equation}
{\cal C} = 2 \Theta_0 + \Theta^1 + \Theta^2 + \Theta^3 + \Theta^4,
\label{eq:r26}
\end{equation}
with $(\Theta_0,\Theta^1)=(\Theta_0,\Theta^2)=(\Theta_0,\Theta^3)=
(\Theta_0,\Theta^4)=1$. In general, the four external points of $D$-diagrams 
can be associated with the four ${\bf Z}_2$ orbifold points of the torus. 

\subsection{Singularities of Type $\hat{A}_{n-1}$.}

We will now consider the case
\begin{equation}
S_{\rho} = \left( \begin{array}{cc} 1 & n \\ 0 & 1 \end{array} \right), 
\label{eq:r27}
\end{equation}
which is of infinite order. A local model for this monodromy can be defined by 
\begin{equation}
\tau(t) = \frac {1}{2 \pi i} n \log t.
\label{eq:r28}
\end{equation}
Using the variable $\rho$ defined in (\ref{eq:r9}), we get, for the group 
${\cal G}_{\rho}$ of automorphisms,
\begin{equation}
g(k,n_1,n_2):(\rho,\lambda) \rightarrow (\rho+k,\lambda+n_1n \rho+n_2),
\label{eq:r29}
\end{equation}
and the local model for the elliptic fibration, out of the singular point,
\begin{equation}
({\cal U}_{\rho} \times {\bf C})/{\cal G}_{\rho},
\label{eq:r30}
\end{equation}
i. e., fibers of the type of elliptic curves, with elliptic modulus $n\rho$. 
A simple way to think about these elliptic curves is in terms of cyclic 
unramified coverings \cite{Fay}. Let us recall that a cyclic unramified covering, 
$\Pi : \hat{C} \rightarrow C$, of order $n$, of a curve $C$ of genus $g$, is 
a curve $\hat{C}$ of genus 
\begin{equation}
\hat{g}=ng+1-n.
\label{eq:r31}
\end{equation}
Thus, for $g=1$, we get $\hat{g}=1$, for arbitrary $n$. Denoting by $\tau$ 
the elliptic modulus of $C$, in case $g=1$, the elliptic modulus of $\hat{C}$ 
is given by
\begin{equation}
\hat{\tau}=n \tau.
\label{eq:r32}
\end{equation}
Moreover, the generators $\hat{\alpha}$ and $\hat{\beta}$ of $H_1(\hat{C};{\bf Z})$ 
are given in terms of the homology basis $\alpha$, $\beta$ of $C$ as
\begin{eqnarray}
\Pi \hat{\alpha} & = & \alpha, \nonumber \\
\Pi \hat{\beta}  & = & n \beta, 
\label{eq:r33}
\end{eqnarray}
with $\Pi$ the projection $\Pi: \hat{C} \rightarrow C$. From (\ref{eq:r32}) 
and (\ref{eq:r29}), we can interpret the elliptic fibration (\ref{eq:r30}) 
as one with elliptic fibers given by $n$-cyclic unramified coverings of a 
curve $C$ with elliptic modulus $\rho$ or, equivalently, $\frac {1}{2 \pi i} 
\log t$. There exits a simple way to define a family of elliptic curves, 
with elliptic modulus given by $\frac {1}{2 \pi i} \log t$, which is the plumbing 
fixture construction. Let $D_0$ be the unit disc around $t=0$, and let 
$C_0$ be the Riemann sphere. Define two local coordinates, $z_a: {\cal U}_a 
\rightarrow D_0$, $z_b : {\cal U}_b \rightarrow D_0$, in disjoint 
neigbourhoods ${\cal U}_a$, ${\cal U}_b$, of two points $P_a$ and $P_b$ of 
${\cal C}_0$. Let us then define
\[
W=\{(p,t) | t \in D_0, p \in C_0-{\cal U}_a - {\cal U}_b, \hbox { or } 
p \in {\cal U}_a, \hbox { with } |z_a(p)|>|t|, \hbox { or } \]
\begin{equation} 
p \in {\cal U}_b, \hbox { with } |z_b(p)|>|t| \},
\end{equation}
and let $S$ be the surface
\begin{equation}
S=\{ xy=t;(x,y,t)\in D_0 \times O_0 \times D_0 \}.
\end{equation}
We define the family of curves through the following identifications
\begin{eqnarray}
(p_a,t) \in W \cap {\cal U}_a \times D_0 & \simeq & (z_a(p_a), \frac {t}{z_a(p_a)}, t) 
\in S, \nonumber \\
(p_b,t) \in W \cap {\cal U}_b \times D_0 & \simeq & (\frac {t}{z_b(p_b)}, 
{z_b(p_b)}, t) \in S.
\end{eqnarray}
For each $t$ we get a genus one curve, and at $t=0$ we get a nodal curve by 
pinching the non zero homology cycles. The pinching region is characterized 
by
\begin{equation}
xy=t,
\label{eq:r34}
\end{equation}
which defines a singularity of type $A_0$. The elliptic modulus of the curves 
is given by
\begin{equation}
\tau(t) = \frac {1}{2\pi i} \log t + C_1 t + C_2,
\label{eq:r35}
\end{equation}
for some constants $C_1$ and $C_2$. We can use an appropiate choice of 
coordinate $t$, such that $C_1=C_2=0$. The singularity at $t=0$ is a singularity 
of type $\hat{A}_0$, in Kodaira's classification, corresponding to
\begin{equation}
S_{\rho} = \left( \begin{array}{cc} 1 & 1 \\ 0 & 1 \end{array} \right).
\label{eq:r36}
\end{equation}
Using now (\ref{eq:r32}) and (\ref{eq:r35}) we get, for the cyclic covering 
of order $n$, the result (\ref{eq:r28}), and the group (\ref{eq:r29}). The 
pinching region of the cyclic unramified covering is given by
\begin{equation}
xy=t^n,
\label{eq:r37}
\end{equation}
instead of (\ref{eq:r34}), i. e., for the surface defining the $A_{n-1}$ 
singularity, ${\bf C}^2/{\bf Z}_n$. Now, we can proceed to the resolution 
of the singularity at $t=0$. The resolution of the singularity (\ref{eq:r37}) 
requires $n-1$ exceptional divisors, $\Theta_1, \ldots, \Theta_{n-1}$. In 
addition, we have the rational curve $\Theta_0$, defined by the complement 
of the node. Thus, we get, at $t=0$,
\begin{equation}
{\cal C} =\Theta_0 + \cdots + \Theta_{n-1},
\label{eq:r38}
\end{equation}
with $(\Theta_0,\Theta_1)=(\Theta_0,\Theta_{n-1})=1$, and $(\Theta_i,
\Theta_{i+1})=1$, which is the $\hat{A}_{n-1}$ Dynkin diagram. The group 
of covering transformations of the $n^{th}$ order cyclic unramified 
covering is ${\bf Z}_n$, and the action over the components (\ref{eq:r38}) 
is given by
\begin{eqnarray}
\Theta_i & \rightarrow & \Theta_{i+1}, \nonumber \\
\Theta_{n-1} & \rightarrow & \Theta_0.
\label{eq:r39}
\end{eqnarray}

\subsection{Singularities of Type $\hat{D}_{n+4}$.}

This case is a combination of the two previous examples. Through the same 
reasoning as above, the group ${\cal G}_{\rho}$ is given, for
\begin{equation}
S_{\rho} = \left( \begin{array}{cc} -1 & -n \\ 0 & -1 \end{array} \right).
\label{eq:r40}
\end{equation}
by 
\begin{equation}
g(k,n_1,n_2) : (\rho,\lambda) \rightarrow (k+\rho, (-1)^k (\lambda +n_1 n \rho +n_2)).
\label{eq:r41}
\end{equation}
Using a new variable $\sigma^2=t$, what we get is a set of irreducible components 
$\Theta_0, \ldots \Theta_{2n}$, with the identifications $\Theta_i 
\rightarrow \Theta_{2n-i}$. In addition, we get the four fixed ${\bf Z}_2$ 
orbifold points described above. The singular fiber is then given by
\begin{equation}
{\cal C} = 2 \Theta_0 + \cdots + 2 \Theta_n + \Theta^1 + \Theta^2 + 
\Theta^3 + \Theta^4,
\label{eq:r42}
\end{equation}
with the intersections of the $\hat{D}_{n+4}$ affine diagram. It is easy to see 
that in this case we also get
\begin{equation}
({\cal C})^2 =0.
\label{eq:r43}
\end{equation}
Defing the genus of the singular fiber by $C^2=2g-2$, we conclude that $g=1$, 
for all singularities of Kodaira type. Notice that for rational singularities, 
characterized by non affine Dynkin diagrams of ADE type \cite{Artin}, we get self intersection 
${\cal C}^2=-2$, which corresponds to genus equal zero.

\section{Decompactification and Affinization.}

The general framework in which we are working in order to get four dimensional 
$N=1$ gauge theories is that of M-theory compactifications on elliptically 
fibered Calabi-Yau fourfolds, in the limit $\hbox {Vol }(E)=0$, with $E$ the elliptic fiber. 
As described above, we can interpolate between $N=2$ supersymmetry in three 
dimensions, and $N=1$ in four dimensions, by changing the radius $R$ 
through 
\begin{equation}
\hbox {Vol }(E) = \frac {1}{R}.
\label{eq:r44}
\end{equation}
The three dimensional limit then corresponds to $\hbox {Vol }(E) \rightarrow 
\infty$, and the four dimensional to $\hbox {Vol } (E) \rightarrow 0$. Now, 
we will work locally around a singular fiber of Kodaira $\hat{A}\hat{D}
\hat{E}$ type. As we know, for the Calabi-Yau fourfold $X$,
\begin{equation}
E \rightarrow X \stackrel{\Pi}{\rightarrow} B,
\label{eq:r45}
\end{equation}
the locus $C$ in $B$, where the fiber is singular, is of codimension one 
in $B$, i. e., of real dimension four. Let us now see what happens to the 
singular fiber in the three dimensional limit. In this case, we have 
$\hbox {Vol }(E)=\infty$. A possible way to represent this phenomenon is 
by simply extracting the point at infinity. In the case of $\hat{A}_{n-1}$ 
singularities, as described in previous subsection, taking out the point at infinity 
corresponds to decompactifying the irreducible component $\Theta_0$, that 
was associated with the curve itself. As was clear in this case, we then pass 
from the affine diagram, $\hat{A}_{n-1}$, to the non affine, $A_{n-1}$. More 
generally, as the elliptic fibration we are considering possesses a global section, 
we can select the irreducible component we are going to decompactify as the one 
intersecting with the basis of the elliptic fibration. When we decompactify, 
in the $\hbox {Vol }(E)=0$ limit, what we are doing, at the level of the 
fiber, is precisely compactifying the extra irreducible component, 
which leads to the affine Dynkin diagram.

\section{M-Theory Instantons and Holomorphic Euler Characteristic.}

Using the results of reference \cite{Wsp} a vertical instanton \cite{Wsp} in a Calabi-Yau 
fourfold, of the type (\ref{eq:r45}), will be defined by a divisor $D$ 
of $X$, such that $\Pi(D)$ is of codimension one in $B$, and with holomorphic 
Euler characteristic
\begin{equation}
\chi(D,{\cal O}_D)=1.
\label{eq:r46}
\end{equation}
It is in case (\ref{eq:r46}) that we have two fermionic zero modes \cite{Wsp}, and we 
can define a superpotential contribution associated to $D$. For $N$, the 
normal bundle to $D$ in $X$, which is locally a complex line bundle on $D$, we 
define the $U(1)$ transformation
\begin{equation}
t \rightarrow e^{i \alpha}t,
\label{eq:r47}
\end{equation}
with $t$ a coordinate of the fiber of $N$. The two fermionic zero modes have 
$U(1)$ charge equal one half. Associated to the divisor $D$, we can define 
a scalar field $\phi_D$ that, together with $\hbox {Vol }(D)$ defines the 
imaginary and real parts of a chiral superfield. Under $U(1)$ rotations 
(\ref{eq:r47}), $\phi_D$ transforms as
\begin{equation}
\phi_D \rightarrow \phi_D + \chi(D) \alpha.
\label{eq:r48}
\end{equation}
In three dimensions, this is precisely the transformation of the dual photon field 
as Goldstone boson \cite{AHW}. However, transformation (\ref{eq:r48}) has perfect sense, 
for vertical instantons, in the four dimensional decompactification limit. 
  
Let us now consider an elliptically fibered Calabi-Yau fourfold, with 
singular fiber of $\hat{A}_{n-1}$ type, over a locus $C$ of codimension 
one in $B$. We will assume that the singular fiber is constant over $C$. Moreover, 
in the geometrical engineering spirit, we will impose
\begin{equation}
h_{1,0}(C)=h_{2,0}(C)=0
\label{eq:r49}
\end{equation}
and, thus, $\Pi_1(C)=0$. This prevents us from having non trivial transformations 
on the fiber by going, on $C$, around closed loops, since all closed loops 
are contractible. In addition, we will assume, based on (\ref{eq:r49}), 
that $C$ is an Enriques surface. After impossing these assumptions, we will 
consider divisors $D_i$, with $i=0, \ldots, n-1$, defined by the fibering over 
$C$, in a trivial way, of the irreducible components $\Theta_i$ of the 
$\hat{A}_{n-1}$ singularity \cite{KV}. Ussing the Todd representation of the holomorphic 
Euler class \cite{Fulton},
\begin{equation}
\chi(D) = \frac {1}{24} \int_{D_i} c_1 (\Theta_i) c_2(C)
\label{eq:r50}
\end{equation}
we get, for $C$ an Enriques surface,
\begin{equation}
\chi(D_i) =1.
\label{eq:r51}
\end{equation}
Interpreting now the $t$ variable (\ref{eq:r47}) on the fiber of the normal 
bundle $N$ of $D$ in $X$ as the $t$ variable used in our previous 
description of Kodaira singularities of type $\hat{A}_{n-1}$, we can derive the 
transformation law, under the ${\bf Z}_n$ subgroup of $U(1)$, of the scalar 
fields $\phi_{D_i}$ associated to these divisors. Namely, from (\ref{eq:r39}) 
we get
\begin{equation}
{\bf Z}_n: \phi_{D_i} \rightarrow \phi_{D_{i+1}},
\label{eq:r52}
\end{equation}
with the ${\bf Z}_n$ transformation being defined by
\begin{equation}
t \rightarrow e^{2 \pi i/n}t.
\label{eq:r53}
\end{equation}
Using now (\ref{eq:r48}), we get
\begin{equation}
{\bf Z}_n : \phi_{D_i} \rightarrow \phi_{D_i} + \frac {2 \pi}{n}.
\label{eq:r54}
\end{equation}
Combining (\ref{eq:r52}) and (\ref{eq:r54}) we get, modulo $2 \pi$,
\begin{equation}
\phi_{D_j} = \frac {2 \pi j}{n} + c,
\label{eq:r55}
\end{equation}
with $j=0,\ldots,n-1$, and $c$ a constant independent of $j$.
  
Let us now consider the divisor ${\cal D}$ obtained by fibering over $C$ the 
singular fiber ${\cal C} = \sum_{j=0}^{n-1} \Theta_j$, defined in (\ref{eq:r38}). In 
this case we need to be careful in order to compute (\ref{eq:r50}). If we 
naively consider the topological sum of components $\Theta_j$ in (\ref{eq:r50}), 
we will get the wrong result $\chi({\cal D})=n$. This result would be correct 
topologically, but not for the holomorphic Euler characteristic we are 
interested in. In fact, what we should write in (\ref{eq:r50}) for 
$\int c_1(\sum_{j=0}^{n-1} \Theta_j)$ is $2(1-g(\sum_{j=0}^{n-1} \Theta_j))$, 
with $g$ the genus of the cycle (\ref{eq:r38}), as defined by 
${\cal C}^2=2g-2$, with ${\cal C}^2$ the self intrsection of the cycle (\ref{eq:r38}) 
which, as for any other Kodaira singularity, is zero. Thus, we get $g=1$, 
and 
\begin{equation}
\chi({\cal D})=0.
\label{eq:r56}
\end{equation}
We can try to intepret the result (\ref{eq:r56}) in terms of the fermionic 
zero modes of each component $\Theta_i$, and the topology of the cycle. In fact, 
associated to each divisor $D_i$ we have, as a consequence of (\ref{eq:r51}), 
two fermionic zero modes. In the case of the $\hat{A}_{n-1}$ singularity, 
we can soak up all zero modes inside the graph, as shown in Figure $1$,


\begin{figure}[ht]
\def\epsfsize#1#2{.6#1}
\centerline{\epsfbox{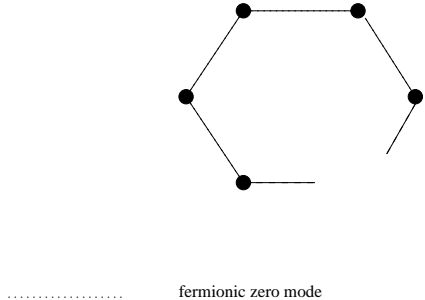}}
\caption{Soaking up of zero modes for $\hat{A}_{n-1}$.}
\end{figure}

 
where from each node, representing one $\Theta_i$, we have two 
fermionic zero mode lines. The soaking up of fermionic zero modes represented 
in the figure is an heuristic interpretation of the result (\ref{eq:r56}).

\section{$\theta$-Parameter and Gaugino Condensates.}

We will, in this section, only consider singularities of
$\hat{A}_{n-1}$ and $\hat{D}_{n+4}$ type. In both cases, and for
each irreducible component $\Theta_i$, we get a divisor $D_i$,
with $\chi(D_i)=1$. Associated to this divisor, we can get a
superpotential term of the order \cite{Wsp}
\begin{equation}
\int d^2 \theta e^{-(V(D_i))+ i \phi_{D_i}}),
\label{eq:r60}
\end{equation}
where $V(D_i)$ means the volume of the divisor $D_i$. As
explained above, we are using vertical instanton divisors $D_i$,
defined by a trivial fibering of $\Theta_i$ over the singular
locus $C \subset B$, satisfying conditions (\ref{eq:r49}). In
order to get the four dimensional $N=1$ limit, we will take the
limit $\hbox {Vol }(E)=\frac {1}{R} \rightarrow 0$. Since the
singular fibers are, topologically, the union of irreducible
components (see (\ref{eq:r38}) and (\ref{eq:r42})), we can write
\begin{equation}
\hbox {Vol }(\Theta_i) = \frac {1}{R \hbox {Cox}},
\label{eq:r61}
\end{equation}
with $\hbox {Cox}$ the Coxeter number of the corresponding
singularity, which equals the total number of irreducible
components. Therefore, we will define $\hbox {Vol }(D_i)$ as
\begin{equation}
\hbox {Vol }(D_i) = \lim_{R \rightarrow \infty} \hbox {Vol }(C)
\frac {1}{R \hbox {Cox}}.
\label{eq:r62}
\end{equation}
If we first consider the $N=2$ supersymmetric three dimensional
theory obtained by compactifying M-theory on the Calabi-Yau
fourfold $X$, i. e., in the limit $R \rightarrow 0$, we know that
only the divisor $\Theta_0$, for the $\hat{A}_{n-1}$ case, is decompactified, passing from the
affine diagram describing an elliptic singularity to the non
affine diagram describing a rational, Artin like, singularity \cite{Artin}. In
that case, the volumes of the $\Theta_i$ components, for $i \neq
0$, are free parameters, corresponding to the Coulomb branch of
the $N=2$ three dimensional theory. In the three dimensional
theory, the factor $\hbox{Vol }(C)$ corresponds to the bare
coupling constant in three dimensions,
\begin{equation}
\hbox {Vol }(C)= \frac {1}{g_3^2},
\label{eq:r63}
\end{equation}
and $\hbox {Vol }(D_i)= \frac {1}{g_3^2} \chi_i$, for $i \neq 0$,
with $\chi_i$ the three dimensional Coulomb branch coordinates.
In the four dimensional case, we must use (\ref{eq:r62}), that
becomes
\begin{equation}
\hbox {Vol }(D_i) = \lim_{R \rightarrow \infty} \frac {1}{g_3^2}
\frac {1}{R \hbox {Cox}} = \frac {1}{g_4^2 \hbox {Cox}}.
\label{eq:r64}
\end{equation}
  
Let us now concentrate on the $\hat{A}_{n-1}$ case, where $\hbox
{Cox}=n$. Using  (\ref{eq:r60}) we get the following
superpotential for each divisor $D_j$,
\begin{equation}
\exp - \left( \frac {1}{g_4^2 {n}} + i \left( \frac {2
\pi j}{ {n}}+ { c } \right) \right).
\label{eq:r65}
\end{equation}
Let us now fix the constant $c$  in (\ref{eq:r65}). In order to do
that, we will use the transformation rules (\ref{eq:r48}). From
the four dimensional point of view, these are the transformation
rules with respect to the $U(1)_R$ symmetry. From (\ref{eq:r48})
we get, that under $t \rightarrow e^{i \alpha}t$,
\begin{equation}
\sum_{i=0} ^{\hbox {n-1}} \phi_{D_i} \rightarrow
\sum_{i=0}^{\hbox {n-1}} \phi_{D_i} + n \alpha.
\label{eq:r66}
\end{equation}
This is precisely the transformation rule under $U(1)_R$ of the
$N=1$ $\theta$-parameter,
\begin{equation}
\theta \rightarrow \theta +n \alpha.
\label{eq:r67}
\end{equation}
In fact, (\ref{eq:r66}) is a direct consequence of the $U(1)$
axial anomaly equation: if we define $\theta$ as
\begin{equation}
\frac {\theta}{32 \pi^2} F \tilde{F},
\label{eq:r69}
\end{equation}
the anomaly for $SU(n)$ is given by
\begin{equation}
\partial_{\mu} j^{\mu}_5 = \frac {n}{16 \pi^2} F
\tilde{F}.
\label{eq:r70}
\end{equation}
The factor $2$, differing (\ref{eq:r69}) from (\ref{eq:r70}),
reflects the fact that we are assigning $U(1)_R$ charge $\frac
{1}{2}$ to the fermionic zero modes. Identifying the $\theta$-parameter 
with the topological sum $\sum_{i=0}^{n-1} \phi_{D_i}$ we get 
that the constant $c$ in (\ref{eq:r65}) is simply
\begin{equation}
\hbox {c} = \frac {\theta}{n},
\label{eq:r71}
\end{equation}
so that we then finally obtain the superpotential
\begin{equation}
\exp - \left( \frac {1}{g_4^2 n} + i \left( \frac {2
\pi j}{n} + \frac {\theta}{n} \right) \right)
\simeq \Lambda^3 e^{2 \pi ij/n} e^{i \theta/n}, 
\label{eq:r72}
\end{equation}
with $j=0, \ldots , n-1$, which is the correct value
for the gaugino condensate.
  
Let us now try to extend the previous argument to the
$\hat{D}_{n+4}$ type of singularities. Defining again the four
dimensional $\theta$-parameter as the topological sum of
$\phi_{D_i}$ for the whole set of irreducible components we get,
for the cycle (\ref{eq:r42}), the transformation rule
\begin{equation}
\theta \rightarrow \theta + \hbox {Cox }. \alpha
\label{eq:h1}
\end{equation}
where now the Coxeter for $\hat{D}_{n+4}$ is $2n+6$. Interpreting
$\hat{D}_{n+4}$ as $O(N)$ gauge groups, with $N=2n+8$, we get
$\hbox {Cox}(\hat{D}_{n+4})=N-2$. Since $\theta$ is defined
modulo $2 \pi$ we get that for $\hat{D}_{n+4}$ singularities the
value of $\phi_{D_i}$, for any irreducible component, is
\begin{equation}
\frac {2 \pi k}{N-2} + \frac {\theta}{N-2},
\label{eq:h2}
\end{equation}
with $k=1, \ldots, N-2$. However, now we do not know how to
associate a value of $k$ to each irreducible component $\Theta_i$
of the $\hat{D}_{n+4}$ diagram. Using (\ref{eq:h2}), we get a set
of $N-2$ different values for the gaugino condensate for $O(N)$
groups:
\begin{equation}
\exp \left( - \frac {1}{g_4^2(N-2)} + i \left( \frac {2 \pi
k}{N-2} + \frac {\theta}{N-2} \right) \right),
\label{eq:h3}
\end{equation}
with $k=1,\ldots,N-2$. However, we still do not know how to
associate to each $\Theta_i$ a particular value of $k$. A
possibility will be associating consecutive values of $k$ to
components with non vanishing intersection; however, the topology
of diagrams of type $D$ prevents us from doing that globally.
Notice that the problem we have is the same sort of puzzle we
find for $O(N)$ gauge groups, concerning the number of values for
$<\lambda \lambda>$, and the value of the Witten index, which in
diagramatic terms is simply the number of nodes of the diagram.
In order to unravel this puzzle, let us consider more closely the
way fermionic zero modes are soaked up on a $\hat{D}_{n+4}$
diagram. We will use the cycle (\ref{eq:r42}); for the components
$\Theta^1$ to $\Theta^4$, associated to the ${\bf Z}_2$ orbifold
points, we get divisors with $\chi=1$. Now, for the components $2
\Theta_0, \ldots, 2 \Theta_n$ we get, from the Todd
representation of the holomorphic Euler characteristic,
\begin{equation}
\chi=4.
\label{eq:h4}
\end{equation}
The reason for this is that the cycle $2 \Theta$, with
$\Theta^2=-2$, has self intersection $-8$. Of course,
(\ref{eq:h4}) refers to the holomorphic Euler characteristic of
the divisor obtained when fibering over $C$ any of the cycles $2
\Theta_i$, with $i=0, \ldots, n$. Equation (\ref{eq:h4}) implies
$8$ fermionic zero modes, with the topology of the soaking up of zero modes of
the $\hat{D}_{n+4}$ diagram, as represented in Figure
$2$.


\begin{figure}[ht]
\def\epsfsize#1#2{.6#1}
\centerline{\epsfbox{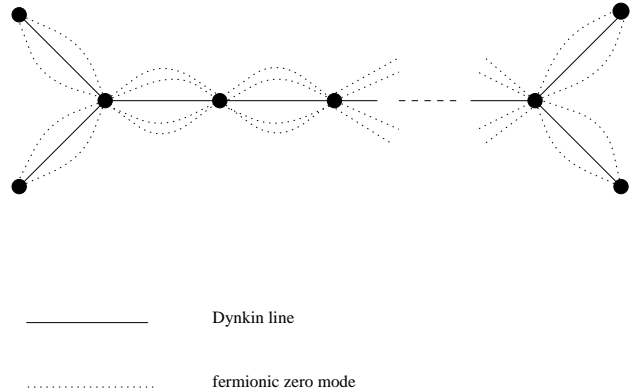}}
\caption{Soaking up of zero modes for $\hat{D}_{n+4}$.}
\end{figure}


 Notice that the contribution to $\chi$ of $2 \Theta$ is different form that 
of $(\Theta_1 + \Theta_2)$, with $(\Theta_1 . \Theta_2)=0$;
namely, for the first case $\chi=4$, and $\chi=2$ for the second.
For the $\hat{D}_{n+4}$ diagram, we can define: $i)$ The Witten
index $\hbox {tr }(-1)^F$, as the number of nodes, i. e., $5+n$;
$ii)$ The Coxeter number, which is the number of irreducible
components, i. e., $2n+6$ and $iii)$ The number of intersections
as represented by the dashed lines in Figure $4$, i. e.,
$8+4n$. From the point of view of the Cartan algebra, used to
define the vacuum configurations in \cite{Wind}, we can only feel
the number of nodes. The $\theta$-parameter is able to feel the
Coxeter number; however, we now find a new structure related to
the intersections of the graph. In the Witten index case, the
nodes corresponding to cycles $2 \Theta_i$, with $i=0, \ldots ,n$
contribute with one, in the number of $<\lambda \lambda>$ values
with two, and in the number of intersections with four. This
value four calls for an orientifold interpretation of these
nodes. The topological definition of the $\theta$-parameter
implicitely implies the split of this orientifold into two
cycles, a phenomena recalling the F-theory description \cite{Sen}
of the Seiberg-Witten splitting \cite{SW}. Assuming this
splitting of the orientifold, the only possible topology for the
soaking up of zero modes is the one represented in Figure $3$,
where the ``splitted orientifold'' inside the box is associated
to four zero modes, corresponding to $\chi=2$ for a cycle
$\Theta_1 + \Theta_2$, with $\Theta_1 . \Theta_2=0$. 


\begin{figure}[ht]
\def\epsfsize#1#2{.6#1}
\centerline{\epsfbox{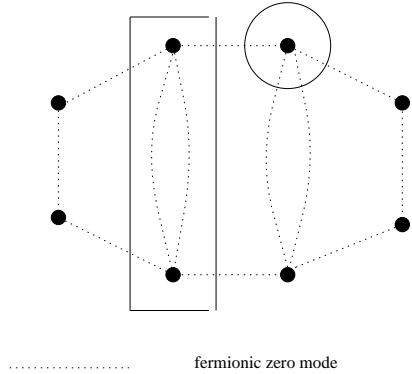}}
\caption{Orientifold splitting.}
\end{figure}


On the
other hand, each node surrounded by a circle in Figure $5$ represents
itself the disconnected sum of two non singular rational curves;
thus, we represent each ``orientifold'' mode by four rational
curves, with the intersections depicted inside the box of Figure
$5$. When we forget about internal lines in Figure $5$, we
get the cyclic ${\bf Z}_{2n+6} \equiv {\bf Z}_{N-2}$ structure of
equation (\ref{eq:h3}). It is clear that much more is necessary
in order to reach a complete description of the $O(N)$ vacuum
structure.

\section{Domain Walls and Intersections.}

The discussion in the previous section already raises the problem
known as $\theta$-puzzle. In fact, and discussing again only the
$SU(n)$ case, the transformation law (\ref{eq:r67}) together with the
very definition fo the $\theta$-angle as the topological sum
$\sum_{i=0}^{n-1} \phi_{D_i}$ would imply that $\theta$ is the
scalar field $\phi_{\cal D}$ of the $6$-cycle associated to the $\hat{A}_{n-1}$ cycle, ${\cal
C}= \sum_{i=0}^{n-1} \Theta_i$. On the basis of (\ref{eq:r48}),
this will be equivalent to saying that $\chi({\cal D})=n$,
instead of zero. This is, in mathematical terms, the
$\theta$-puzzle. The mathematical solution comes from the fact
that $\chi({\cal D})=0$. In this section we will relate this
result, on the value of the holomorphic Euler chareacteristic, to
the appearance of domain walls \cite{ds,kss}. To start
with, let us consider a cycle ${\cal C} = \Theta_1 + \Theta_2$,
with $(\Theta_1 . \Theta_2)=1$. The self intersection can be
expressed as
\begin{equation}
({\cal C}. {\cal C})=-2-2+2,
\label{eq:r76}
\end{equation}
where the $-2$ contributions come from $\Theta_1^2$ and
$\Theta_2^2$, and the $+2$ comes from the intersection between
$\Theta_1$ and $\Theta_2$. As usual, we can consider ${\cal C}$ 
trivially fibered on an Enriques surface. The holomorphic Euler
chracteristic of the corresponding six cycle can be written as 
\begin{equation}
\chi = \frac {1}{2} (- {\cal C}^2).
\label{eq:h5}
\end{equation}
Using now the decomposition (\ref{eq:r76}) we get two
contributions of one, coming from the components $\Theta_1$ and
$\Theta_2$, considered independently, and a contribution of $-1$
from the intersection term $+2$ in (\ref{eq:r76}). In this sense,
the intersection term can be associated to two fermionic zero
modes, and net change of chiral charge oposite to that of the
$\Theta_i$ components. When we do this for the cycle ${\cal C}$
of $\hat{A}_{n-1}$ singularities, we get that each intersection
is soaking up two zero modes, leading to the result that
$\chi({\cal C})=0$. A graphical way to represent equation
(\ref{eq:r76}) is presented in Figure $4$.


\begin{figure}[ht]
\def\epsfsize#1#2{.6#1}
\centerline{\epsfbox{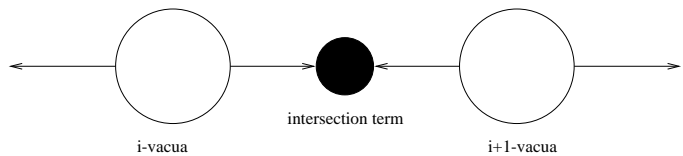}}
\caption{Intersection term.}
\end{figure}


Now, we will wonder about the physical interpretation of the
intersection terms leading to $\chi({\cal C})=0$ for all Kodaira
singularities. The simplest, and most natural answer, is
certainly domain walls extending between different vacua, or
values of $<\lambda \lambda>$.

From the point of view of zero mode counting, the ``intersection
term'' behaves effectively as an anti-instanton with two
fermionic zero modes. One of these fermionic zero modes, let us
say $\psi_{j,j+1}$, is associated to the intersection of
$\Theta_{j}$ with $\Theta_{j+1}$ and the other $\psi_{j+1,j}$
with the intersection of $\Theta_{j}$ and $\Theta_{j+1}$. Thus, extending naively the 
computation done for irreducible components, the contribution of
the black box in Figure $4$ should be of the order
\begin{equation}
\Lambda^3 e^{2 \pi ij/n} (1-e^{2 \pi i/n}).
\label{eq:h10}
\end{equation}
In result (\ref{eq:h10}), interpreted as the contribution of the
intersection term, the most surprising fact is the appearance of
$\Lambda^3$, since now we are geometrically considering simply a
point; the factor $\Lambda^3$ in the computation of the gaugino
condensate comes from the volume of the divisor. In the same way
as we interpret M-theory instantons as fivebranes wrapped on the
six-cycles used to define the instanton, we can think of the
intersection terms as fivebranes wrapped on the cycle $C \times
\{(\Theta_i . \Theta_{i+1})\}$, i. e., the product of the singular
locus $C$ and the intersection point. The fivebrane wrapped on
this cycle defines, in four dimensions, a domain wall, let us say
interwining between the vacua $i$, at $x_3=+ \infty$, and the
vacua $i+1$, at $x_3=- \infty$, where the coordinate $x_3$ is
identified with the unwrapped direction. It is in this sense that
we should use (\ref{eq:h10}) to define the energy density, or
tension, of the domain wall. In the four dimensional limit,
$\hbox {Vol }(E)$ goes zero as $\frac {1}{R}$; moreover, the
local engineering approach works in the limit where the volume of
the singular locus $C$ is very large, so that we can very likely
assume that intersection terms behave like (\ref{eq:h10}), with
$\Lambda^3$, but only in the four dimensional limit. Cyclicity of
the $\hat{A}_{n-1}$ diagram allows us to pass from the $j$ to the
$j+1$ vacua in two different ways: $n-1$ steps, or a single one.
The sum of both contributions should define the physical domain
wall; thus the energy density will behave as
\begin{equation}
n \Lambda^3 |e^{2 \pi ij/n} (1-e^{2 \pi i/n})|.
\label{eq:h11}
\end{equation}
  
The extension of the previous argument to the case of $O(N)$
groups is certainly more involved, due to the topology of the
$\hat{D}$ diagram, and the presence of orientifolds. It would
certainly be interesting studying the interplay between
orientifolds and domain walls in this case.
  
Finally, we will say some words on the QCD string. In reference
\cite{Wbqcd}, the geometry of QCD strings is intimately related
to the topological fact that
\begin{equation}
H_1(Y/Z;{\bf Z}) = {\bf Z}_n,
\label{eq:r77}
\end{equation}
where $\Sigma$ is a rational curve associated to the
configuration of fourbranes, and $Y=S^1 \times {\bf R}^5$ is the
ambient space where $\Sigma$ is embedded (see \cite{Wbqcd} for
details). The QCD string is then associated to a partially
wrapped membrane on a non trivial element of $H_1(Y/\Sigma;{\bf
Z})$. Recall that $H_1(Y/\Sigma;{\bf Z})$ is defined by
one-cycles in $Y$, with boundary on $\Sigma$. The previous
discussion was done for $SU(N)$ gauge groups. Using our model of
$\hat{A}_{n-1}$ singularities, described in section 2, the
analog in our framework of (\ref{eq:r77}) is equation
(\ref{eq:r33}). Then we can, in the same spirit as in reference
\cite{Wbqcd}, associate the QCD string to paths going from $p_k$
to $p_{k+1}$, where $p_k$ are the intersection points,
\begin{equation}
\Theta_k . \Theta_{k-1} = p_k.
\label{eq:r78}
\end{equation}
Geometrically, it is clear that the tension of this QCD string is
the square root of the domain wall tension. By construction, the
QCD string we are suggesting here ends on domain walls, i. e., on
intersection points.

To end up, let us include some comments on the existence of
extra vacua, as suggested in \cite{KS}. It is known
that the strong coupling computation of $<\lambda \lambda>$ does
not coincide with the weak coupling computation; more precisely\cite{NSVZ2},
\begin{equation}
<\lambda \lambda>_{\hbox {sc}} < \: <\lambda \lambda>_{\hbox
{wc}}.
\label{eq:r75}
\end{equation}
In the framework of M-theory instanton computations, the
numerical factors will depend in particular on the moduli of
complex structures
of the Calabi-Yau fourfold. In the strong coupling regime we must consider
structures preserving the elliptic fibration structure and the
Picard lattice. In the weak coupling regime, where the
compuatation is performed in the Higgs phase, the amount of
allowed complex structures contributing to the value of $<\lambda
\lambda>$ is presumably larger. Obviously, the previous argument is only
suggesting a possible way out of the puzzle (\ref{eq:r75}).

Equally, at a very speculative level, the extra vacua, with no
chiral symmetry breaking, could be associated to the cycle ${\cal
D}$ defining the singular fiber, a cycle that we know leads to
$\chi=0$, and therefore does not produce any gaugino condensate.
Notice that any other cycle with $\chi \neq 0$ will lead, if
clustering is used, to some non vanishing gaugino condensates, so
that ${\cal D}$ with $\chi=0$ looks like a possible
candidate to the extra vacua suggested in \cite{KS}. If this
argument is correct this extra vacua will appears for any Kodaira
singularity i. e. in any ADE $N=1$ four dimensional gauge theory.

It is important to stress that the $\theta$-puzzle is not
exclusive of $N=1$ gluodynamics. In the $N=0$ case the
Witten-Veneziano formula \cite{Wi,Ve} for the $\eta'$ mass 
also indicates a dependence of the vacuum 
energy on $\theta$ in terms of $\frac {\theta}{N}$, 
which means a set of entangled "vacuum" states. 
In our approach to $N=1$ the origin of this entanglement is due to the fact 
that $\chi=0$ for the singular cycle. In fact, $\chi({\cal D})=0$ means that the set 
of divisors $D_i$, plus the intersections, i. e., the domain walls, are 
invariant under $U(1)$, as implied by equation (\ref{eq:r48}). If 
we naively think of something similar in $N=0$
and we look for the origin of vacuum entanglement in intersections we maybe should think in
translating the topology of intersections into topological properties 
of abelian proyection gauges \cite{thooft}.

\vspace{20 mm}
  
\begin{center}
{\bf Acknowledgments}
\end{center}

The author would like to thank U. Bruzzo , C. Reina and R. Hernandez for discussions and
to U. Bruzzo for bringing my attention to reference \cite{Fulton}.
This work is partially supported by European Community grant 
ERBFMRXCT960012, by grant AEN96-1655.

\end{document}